\documentstyle[prl,aps,floats,epsf,color]{revtex}
\begin{document}
\baselineskip=12pt
\def\be{\begin{equation}}
\def\ee{\end{equation}}
\def\bea{\begin{eqnarray}}
\def\eea{\end{eqnarray}}
\def\E{{\rm e}}
\def\bearst{\begin{eqnarray*}}
\def\eearst{\end{eqnarray*}}
\def\peleven{\parbox{11cm}}
\def\peffec{\peight{\bearst\eearst}\hfill\peleven}
\def\pspace{\peight{\bearst\eearst}\hfill}
\def\ptwelve{\parbox{12cm}}
\def\peight{\parbox{8mm}}
\twocolumn[\hsize\textwidth\columnwidth\hsize\csname@twocolumnfalse\endcsname

\title
{ Intermittency of Height Fluctuations in  Stationary State of
The Kardar-Parisi-Zhang Equation with infinitesimal surface
tension in $1+1$ Dimensions }
\author
{ S.M.A. Tabei $^a$, A. Bahraminasab $^a$,
 A. A. Masoudi $^c$, S.S. Mousavi $^a$ \\ and M. Reza Rahimi
Tabar $^{a,b}$ }
\address
{\it $^a$  Department of Physics, Sharif University of
Technology, P.O. Box 11365-9161, Tehran, Iran \\
$^b$ CNRS UMR 6529, Observatoire de la C$\hat o$te d'Azur, BP
4229, 06304 Nice Cedex 4, France \\
 $^c$  Department of Physics, Alzahra University, Tehran, 19834,
 Iran}

%rahimi@netware2.ipm.ac.ir \\}

%\date{01/02/98}
\maketitle

%\date{00/07/2000}
%\maketitle

%%%%%%%%%%%%%%%%%%%%%%%%%%%%%%%%%%%%%%%%%%%%%%%%%%%%%%
%ABSTRACT
%%%%%%%%%%%%%%%%%%%%%%%%%%%%%%%%%%%%%%%%%%%%%%%%%%%%%%
\begin{abstract}

The Kardar-Parisi-Zhang (KPZ) equation with infinitesimal surface
tension, dynamically develops sharply connected valley structures
within which the height derivative is not continuous. We discuss
the intermittency issue in the problem of stationary state forced
KPZ equation in $1+1$--dimensions. It is proved that the moments
of height increments  $C_a = < | h (x_1) - h (x_2) |^a
> $ behave as $ |x_1 -x_2|^{\xi_a}$ with $\xi_a = a$
 for length scales $|x_1-x_2| << \sigma$. The length scale $\sigma$ is the characteristic length of the
forcing term. We have checked the analytical results by direct
numerical simulation.

\end{abstract}
\hspace{.3in}
\newpage
]
\section{Introduction}
 The growth, formation and morphology of interfaces has
been one of the recent interesting fields of study because of its
high technical and rich theoretical advantages.
 On account of the disorder
nature embedded in the surface growth, stochastic differential
equations have been used as a suitable tool for understanding the
behavior of various growth processes. Such equations typically
describe the interfaces at large length scales, which means that
the short length scale details has been neglected in order to
derive a continuum equation by focusing on the coarse grained
properties. A great deal of recent theoretical modeling has been
started with the work of Edward and Wilkinson  describing the
dynamics of height fluctuations by a simple linear stochastic
equation [1-5]. By adding a new term proportional to the square of
the height gradient, Kardar, Parisi and Zhang made an appropriate
description for lateral interface growth [6].
 The $(1+1)$-dimensional
forced KPZ equation is written as
\begin{equation}\label{1}
 h_t(x,t) \nonumber\\
 = \frac{\alpha}{2}
 (h_{x})^2+\nu h_{xx}+f(x,t)
\end{equation}
where $\alpha\geq0$ and $f(x,t)$ is a zero-mean, statistically
homogeneous, white in time random force with covariance
\begin{equation}
\langle f(x,t)f(x',t')\rangle=2D_0D(x-x')\delta(t-t').
\end{equation}

Typically the spatial correlation of the forcing is considered to
be a delta function, mimicking the short length correlation. Here
the spatial correlation is considered as
\begin{equation}\label{}
D({x-x'})=\frac{1}{({\pi\sigma^2})^{1/2}}\exp(-\frac{(
{x-x'})^2}{\sigma^2})
\end{equation}
where $\sigma$ is  much less than the system size $L$, i.e.
$\sigma << L$, which  represents a short range character for the
random forcing. The KPZ equation has come to famous the "Ising
model" of non-equilibrium physics. It is indeed the simplest
equation nevertheless capturing the main determinants of the
growth dynamics i.e. nonlinearity, stochasticity, and locality.
The theoretical richness of the KPZ model is partly due to close
relationships with other areas of statistical physics. It is shown
that there is a mapping between the equilibrium statistical
mechanics of a two dimensional smectic-A liquid crystal onto the
non-equilibrium dynamics of the (1+1)- dimensional stochastic KPZ
equation [7]. It has been shown in [8] that, one can map the
kinetics of the annihilation process $A + B \rightarrow 0$ with
driven diffusion onto the (1+1)-dimensional KPZ equation. Also
the KPZ equation is closely related to the dynamics of a
sine-Gordon chain [9], the driven-diffusion equation [10,11], high
$T_c$- superconductor [12] and directed paths in the random media
[13-26] and  charge density waves [27], dislocations in
disordered solids [3], formation of large-scale structure in the
universe [28-31] , Burgers turbulence [32-60] and etc.

It is useful to rescale the KPZ equation as $h'=h/h_0$, ${\bf
r'=r/r_0}$ and $t'=t/t_0$. If we let $h_0=(\frac{D_0}{\nu
})^{1/2}$ and $t_0= \frac{r_0 ^2}\nu $, where $r_0$ is a
characteristic length, all of the parameters can be eliminated,
except the coupling constant $g= \frac{\alpha ^2D_0}{\nu ^3}$.
The limit $g\rightarrow \infty $ (or zero tension limit, $\nu
\rightarrow 0$), is known as the strong coupling limit [60]. Phase
diagram information extracted from the renormalisation group flow
indicates that $d=2$ plays the role of a lower critical
dimension. For $d \leq 2$, the Gaussian fixed point ($\alpha = 0)
$ is infrared-unstable, and there is a crossover to the stable
strong coupling fixed point. For $d > 2$, a third fixed point
exists, which represents the roughening transition. It is
unstable and appears between the Gaussian and strong coupling
fixed points which are now both stable. Only the critical indices
of the strong-coupling regime ($g \rightarrow \infty $ or $\nu
\rightarrow 0$) are known in 1+1 dimensions and their values in
higher dimensions as well as properties of the roughening
transition have been known only numerically [61-67], and the
various approximation schemes [68-76].

For finite $\sigma$, in the strong coupling limit ($\nu
\rightarrow 0$) nonlinear term in the KPZ equation will dominate.
The nonlinearity of the KPZ equation in this limit includes the
possibility of singularity formations in a finite time as a
result of the local minima instability.  Meaning that there is a
competition between the diffusion smoothing effect ( the
Laplacian term), and the enhancement of non-zero slopes. In one
spatial dimension the sharp valleys are developed in a finite
time. As indicated in figs.(1) and (2), the geometrical picture
consists of a collection of sharp valleys intervening a series of
hills in the stationary state [77].

The main difficulty with the KPZ equation is that it is
controlled, in all dimensions, by a strong disorder ( or strong
coupling) fixed point and efficient tools are missing to
calculate the exponents and other universal properties e.g.
scaling functions, amplitudes, etc. Despite the fact that in one
dimension, the exponents are known, but many properties,
including the probability density function (PDF) of the height of
a growing interface have been so far measured only in numerical
simulations.

In this article, the statistical properties of the KPZ equation
in the strong coupling limit ($\nu \rightarrow 0$) is
investigated. The limit is singular, i.e. through which the
surface develops sharp valleys. So starting with a flat surface
after a finite time scale, $t_c \simeq (\pi)^{1/6} {D_0}^{-1}
\alpha^{-2/3} \sigma^{5/3} $ [77], sharp valley singularities are
dynamically developed. In the singular points spatial derivative
of the field $h(x,t)$ is not continuous. One of the main problem
in this area is the scaling behavior of moments of height
increments
 $C_a = < |h(x_1) - h(x_2)|^a >$ and the probability density
function (PDF) of $ \delta h = h(x_1) - h(x_2) $, that is
$P(\delta h)$.  Inspired by the methods proposed recently  by
Weinan E and Vanden Eijnden [47], a statistical method is
developed to describe the moments of the height and height
gradient increments.
 We derive a master equation for joint PDF of the height and its gradient
 increments in 1+1 dimensions. It is shown that in the stationary state where the
sharp valleys are fully developed, the relaxation term with
infinitesimal surface tension leading to an unclosed term  in the
PDF's equation. However we show that the unclosed term can be
expressed in terms of statistics of some quantities defined on
the sharp valleys. We identify each sharp valley in position
$y_0$ with three quantities, namely the gradients of $h$ in the
positions $y_{0^+}$, $ y_{0^-}$ and its height from the $\bar h$.
The dynamics of these quantities are given in [77].
 Here it is proved
that to leading order, when $|x_1 - x_2| << \sigma$, fluctuation
of the height field is not intermittent. The analytic form of the
amplitudes of the structure functions is also given. The absence
of the intermittency means that $C_a = <|h(x_1) - h(x_2)|^a>$
scales as $|x_1 - x_2|^{\xi_a}$, where $\xi_a$ is a linear
function of $a$. It is proved that for length scales $|x_1 - x_2|
<< \sigma$, the exponents $\xi_a$ are equal to $a$.

The paper is organized as follows; in section two, the known
results for the moments of height increments $C_a = <|h(x_1) -
h(x_2)|^a>$ ,for length scales $|x_1-x_2| >> \sigma$,  are
expressed. In section three, we derive the master equation for
the joint PDF of height and its gradient increments for given
surface tension  $\nu$ and for length scales $|x_1-x_2| <<
\sigma$. It is shown that the surface tension term makes the
PDF's equation unclosed.
 In section 4 we will consider the
limit of $\nu \rightarrow 0$ of the master equation and derive
the scaling exponents of height increments moments. Also a
comparison between the analytical results and direct numerical
simulation are given. Details of calculations are presented in the
appendices A and B.

\section{ Scaling exponents of height-difference moments for forced KPZ equation and for the length
scales $|x_1 - x_2| >> \sigma. $ }

 In this section a review of the known results for the scaling exponent of
height increments moments for the KPZ equation in
$1+1$-dimensions with white in time and space forcing is given.
Indeed the limit $\sigma \rightarrow 0$ is considered in equation
(2). In this limit the equation (2) can be written as follows;
\begin{equation}
\langle f(x,t)f(x'.t')\rangle = 2D_0\delta(x-x')\delta(t-t')
\end{equation}

For this type of forcing, $\Pi[\widetilde{h}(x,t)]$, the
probability functional of $\widetilde{h}(x,t)=h(x,t)-\langle h
\rangle$ satisfies the functional Focker-Planck equation [1,2],
\begin{eqnarray}\label{focker planck}
\frac{\partial}{\partial t}\Pi &&= \int d^d
x\frac{\delta}{\delta\widetilde{h}(x)} [\big( \frac{\alpha}{2}
(\nabla h)^2+\nu \nabla^2h\big)\Pi] \cr \nonumber \\
&&+ D_0\int d^d x\frac{\delta^2}{\delta\widetilde{h}^2(x)}\Pi,
\end{eqnarray}
where its solution in the (1+1)-dimensions is
 \bea
\Pi=\exp \big[-\frac{\nu}{2D_0}\int dx(h_x)^2\big].
 \eea

Therefore if one introduce $G(x-x')=\langle \widetilde{h}(x)
\widetilde{ h}(x')\rangle$ as a Green's function, then it
satisfies the following differential equation
\begin{eqnarray}
\partial_{xx}G(x-x')=-\frac{D_0}{\nu}\delta(x-x')\nonumber\\
\end{eqnarray}
so that $\langle \widetilde{h}(x) \widetilde{
h}(0)\rangle=-\frac{D_0}{\nu}|x|$.
 Now we can write
the second moment of height increments for small $x$'s as
\begin{eqnarray}
\langle|h(x)-h(0)|^2\rangle = \frac{2D_0}{\nu}| x|.
\end{eqnarray}

In a similar way it can be seen that the higher moments,
$\langle|h(x)-h(0)|^a \rangle$ scale with $ x$ as
$|x|^{\frac{a}{2}}$, which means that for the length scale $
\sigma << |x_1 -x_2| << L$, the exponents are $\xi_a =
\frac{a}{2}$.

There are a few comments on
 the result obtained for the
functional PDF,  equation (6).
 It is evident that the probability density functional (in
$1+1$-dimensions ) is independent of the coefficient of the
nonlinear term i.e. $\alpha$, so the result is independent of the
strength of the coupling constant. As it can be seen the scaling
relation is similar to an ordinary random walk problem. If one
considers the random force with smooth spatial correlation, the
problem changes to a more complicated one and there is no any
closed solution for the functional PDF. In the next sections we
will show that the moments of the height increments for the length
scales $ |x_1 -x_2| << \sigma $ has the scaling exponents
 $\xi_a = a$ and the amplitude of the moments are depend on the
coefficient of the nonlinear term $\alpha$.

\section{The master equation governing the probability density function of the
height--difference and gradient--difference for given surface
tension}

In this section , focusing on the (1+1)-dimensional  KPZ
equation  and it's corresponding Burgers equation, the master
equation describing the evolution of the joint two point PDF,
$P(h(x_1)-h(x_2), u(x_1)-u(x_2))$ of the height and corresponding
height gradients increments is derived. The (1+1) dimensional KPZ
equation is written as
\begin{equation}\label{1}
 h_t(x,t) \nonumber\\
 = \frac{\alpha}{2}
 h_{x}^2+\nu h_{xx}+f(x,t)
\end{equation}
where $\alpha\geq0$ and $f(x,t)$ is a zero-mean, statistically
homogeneous, white in time random force. Its covariance is given
by eq.(2). Using the map  $-\partial_x h=u$, the corresponding
Burgers equation is written as
\begin{eqnarray}\label{burgers}
u_{t}=-\alpha uu_{x}+\nu u_{xx}-f_{x}(x,t).  %\nonumber\\
\end{eqnarray}

 Defining the two point generating function as $
Z(\lambda_{1},\lambda_{2},\mu_{1},\mu_{2},x_1,x_2,t)=\langle\Theta\rangle$,
where $\Theta$ is defined as,
\begin{eqnarray}
\Theta:=\exp{(-i\lambda h_1-i\lambda{h}_2 -i\mu_1u_1-i\mu_2u_2)}.
\end{eqnarray}

 The
fields $h_1$ and $h_2$ are  the height of the surface at points
$x_1$ and $x_2$. The fields $u_1= -\partial_{x_1} h_1$ and $u_2=
-\partial_{ x_2}h_2$ are related to the corresponding height
gradients.  As it is seen the generating function is the ensemble
average of $\Theta$.
 The time evolution of $Z$ will be
\begin{eqnarray}\label{evolut}
Z_{t} &=& -i\lambda_{1}\langle h_{1t}\Theta\rangle
-i\lambda_{2}\langle h_{2t}\Theta\rangle \cr \nonumber \\
&-& i\mu_{1}\langle u_{1t}\Theta\rangle-i\mu_{2}\langle
u_{2t}\Theta\rangle.
\end{eqnarray}

Using the equations (\ref{1}) and (\ref{burgers}) and noting that,
in  equation (12), $h_1$, $h_2$, $u_1$ and $u_2$ can be
substituted by $h_1 \rightarrow
 i\frac{\partial}{\partial\lambda_1}$, $h_2 \rightarrow
 i\frac{\partial}{\partial\lambda_2}$, $u_1 \rightarrow
 i\frac{\partial}{\partial\mu_1}$ and $u_2 \rightarrow
 i\frac{\partial}{\partial\mu_2
 }$, the time evolution of $Z$ can be rewritten as
\begin{eqnarray}\label{evolut}
Z_{t}&=& i\frac{\alpha\lambda_{1}}{2}\langle
\Theta\rangle_{\mu_1\mu_1}+i\frac{\alpha\lambda_{2}}{2}\langle
\Theta\rangle_{\mu_2\mu_2} \cr \nonumber \\
&-&\alpha\mu_1\frac{\partial}{\partial\mu_1}\langle
u_{1x_1}\Theta\rangle-\alpha\mu_2\frac{\partial}{\partial\mu_2}\langle
u_{2x_2}\Theta\rangle \cr \nonumber\\
&-& i\lambda_1\langle f_1\Theta\rangle-i\lambda_2\langle
f_2\Theta\rangle +i\mu_1\langle f_{1x_1}\Theta\rangle
+i\mu_2\langle f_{2x_2}\Theta\rangle
\cr \nonumber \\
&+& i\lambda_1\nu\langle u_{1x_1}\Theta\rangle
+i\lambda_2\nu\langle u_{2x_2}\Theta\rangle-i\mu_1\nu\langle
u_{1xx}\Theta\rangle \cr \nonumber \\ &-& i\mu_2\nu\langle
u_{2xx}\Theta\rangle.
\end{eqnarray}

Now using ,
\begin{eqnarray}
\langle
u_{jx}\Theta\rangle=\frac{i}{\mu_j}{\langle\Theta\rangle_{x_j}+
\frac{i\lambda_j}{\mu_j}\langle\Theta\rangle_{\mu_j}},\hspace{1cm}j=1,2
\end{eqnarray}
  the equation governing $Z$ can be written as
\begin{eqnarray}
\label{ZFG} Z_{t} &=& i\frac{\alpha\lambda_{1}}{2}\langle
\Theta\rangle_{\mu_1\mu_1}+i\frac{\alpha\lambda_{2}}{2}\langle
\Theta\rangle_{\mu_2\mu_2} \cr \nonumber \\
&-&\alpha\mu_1\frac{\partial}{\partial\mu_1}(
\frac{i}{\mu_1}{\langle\Theta\rangle_{x_1}
+\lambda_1\langle\Theta\rangle_{\mu_1}}) \cr \nonumber \\
&-&\alpha\mu_2\frac{\partial}{\partial\mu_2}
(\frac{i}{\mu_2}\langle\Theta\rangle_{x_2}+\lambda_2\langle\Theta\rangle_{\mu_2})
\cr \nonumber \\ &+& i\lambda_1\nu (
\frac{i}{\mu_1}{\langle\Theta\rangle_{x_1}+\lambda_1\langle\Theta\rangle_{\mu_1}})
\cr \nonumber \\
&+&i\lambda_2\nu\langle\frac{i}{\mu_2}(\Theta\rangle_{x_2}+\lambda_2\langle\Theta\rangle_{\mu_2})+{\cal
F}+{\cal G}.
\end{eqnarray}

 Here ${\cal F}$ and ${\cal G}$ stand for
\begin{eqnarray}\label{fg}
{\cal F}&=&-i\lambda_1\langle f_1\Theta\rangle-i\lambda_2\langle
f_2\Theta\rangle \cr \nonumber \\  &+& i\mu_1\langle
f_{1x_1}\Theta\rangle+i\mu_2\langle f_{2x_2}\Theta\rangle \cr
\nonumber \\ {\cal G}&=&i\mu_1\nu\langle u_{1xx}\Theta\rangle+
i\mu_2\nu\langle u_{2xx}\Theta\rangle.
\end{eqnarray}

In  equation (\ref{ZFG}) the terms ${\cal F}$ and ${\cal G}$ are
the only terms which are not closed respect to $Z$. Indeed the
term ${\cal F}$ can be
 also closed according to Novikov's theorem ,
\begin{eqnarray}\label{F}
{\cal
F}&=&\big(-(\lambda_1^2+\lambda_2^2)K(0)-2\lambda_1\lambda_2K(x)\big)Z
\cr \nonumber \\
&+& \big(-(\mu_1^2+\mu_2^2)K(0)-2\mu_1\mu_2K(x)\big)Z
\end{eqnarray}
where $K(x)=2D_0D(x)$ and $x = x_1 -x_2$.
 So $G$ is the only term preventing
equation (\ref{ZFG}) to be closed which can be referred to a sort
of dissipative anomaly.

The PDF $P(h_1,h_2,u_1,u_2,x_1,x_2,t)$ is defined as the two-point
joint probability density function (PDF) at the points $x_1$ and
$x_2$ with their related heights $h_1$ and $h_2$, and their
gradients $u_1$ and $u_2$. The PDF can be constructed by Fourier
transforming the generating function $Z$
\begin{eqnarray}
&&P(h_1,h_2,u_1,u_2,x_1,x_2,t)=
\int\frac{d\lambda_1}{2\pi}\frac{d\lambda_2}{2\pi}\frac{d\mu_1}{2\pi}\frac{d\mu_2}{2\pi}\nonumber\\
&&\times\exp{(i\lambda h_1+i\lambda{h}_2
+i\mu_1u_1+i\mu_2u_2)}Z(\lambda_{1},\lambda_{2},\mu_{1},\mu_{2},x_1,x_2,t)\nonumber.
\end{eqnarray}

Fourier transformation  of equation (\ref{ZFG}) gives the
following equation for the PDF
\begin{eqnarray}
\label{Pmaster} -\frac{\partial}{\partial
u_1}\frac{\partial}{\partial u_2}P_t & =&
-\frac{\alpha}{2}\frac{\partial}{\partial
h_1}\frac{\partial}{\partial u_1}\frac{\partial}{\partial
u_2}(u_1^2P) \cr \nonumber \\
&-&\frac{\alpha}{2}\frac{\partial}{\partial
h_2}\frac{\partial}{\partial u_1}\frac{\partial}{\partial
u_2}(u_2^2P) \cr \nonumber\\
&-&\alpha\frac{\partial}{\partial
u_2}P_x+\alpha\frac{\partial}{\partial u_1}P_x \cr \nonumber \\
&-&\alpha\frac{\partial}{\partial h_1}\frac{\partial}{\partial
u_2}(u_1P)\cr \nonumber \\ &-&\alpha\frac{\partial}{\partial
h_2}\frac{\partial}{\partial u_1}(u_2P) \cr \nonumber \\
&-& \alpha\frac{\partial}{\partial u_1}\frac{\partial}{\partial
u_2}(u_1P_x) \cr \nonumber \\ &+& \alpha\frac{\partial}{\partial
u_1}\frac{\partial}{\partial u_2}(u_2P_x)\cr \nonumber \\&+&{\cal
L}(\mu_1\mu_2{\cal F})+{\cal L}(\mu_1\mu_2{\cal G})
\end{eqnarray}
 where $x=x_2-x_1$, $y=\frac{x_1+x_2}{2}$, $\partial_{x_1}=-\partial_{x}+\frac{1}{2}\partial_{y}$ and
 $\partial_{x_2}=\partial_{x}+\frac{1}{2}\partial_{y}$.
The terms  ${\cal L}(\mu_1\mu_2{\cal F})$ and ${\cal
L}(\mu_1\mu_2{\cal G})$  are the Fourier transformations of
equations (16) and (17), multiplied in $\mu_1$ and $\mu_2$, where
for ${\cal L}(\mu_1\mu_2{\cal F})$ is
\begin{eqnarray}
{\cal L}(\mu_1\mu_2{\cal F})&=&-k(0)\frac{\partial}{\partial
u_1}\frac{\partial}{\partial u_2}(\frac{\partial^2}{\partial
h_1^2} + \frac{\partial^2}{\partial h_2^2})P \cr \nonumber
\\ &-& 2k(x)\frac{\partial}{\partial u_1}\frac{\partial}{\partial
u_2}\frac{\partial}{\partial h_1}\frac{\partial}{\partial h_2}P
\cr \nonumber \\ &-& k_{xx}(0)\frac{\partial}{\partial
u_1}\frac{\partial}{\partial u_2}(\frac{\partial^2}{\partial
u_1^2}+ \frac{\partial^2}{\partial u_2^2})P\cr \nonumber \\
&-& 2k_{xx}(x)\frac{\partial^2}{\partial
u_1^2}\frac{\partial^2}{\partial u_2^2}P
\end{eqnarray}
and  ${\cal L}(\mu_1\mu_2{\cal G})$ is defined as
\begin{eqnarray}
{\cal L}(\mu_1\mu_2{\cal G}) &:=& -\nu\big\{ \langle
u_{1x_1x_1}|h_1,h_2,u_1,u_2,x\rangle P\big\}_{u_1u_1u_2} \cr
\nonumber \\ &-& \nu\big\{ \langle
u_{2x_2x_2}|h_1,h_2,u_1,u_2,x\rangle P\big\}_{u_1u_2u_2}.
\end{eqnarray}

For later use we define $G$ as
\begin{eqnarray}\label{G}
G:=G(h_1,h_2,u_1,u_2,x,t)+G(h_1,h_2,u_1,u_2,-x,t)
\end{eqnarray}
where
\begin{eqnarray}\label{G}
G(h_1,u_1,u_1,u_2,x,t)=-\nu \big\{ \langle
u_{1x_1x_1}|h_1,h_2,u_1,u_2,x\rangle P \big \}_{u_1}. \nonumber
\end{eqnarray}

Also we can simply substitute ${\cal L}(\mu_1\mu_2{\cal G})$ with
$G_{u_1u_2}$. In eq.(20), $\langle
u_{1x_1x_1}|h_1,h_2,u_1,u_2,x\rangle$ and $\langle
u_{2x_2x_2}|h_1,h_2,u_1,u_2,x\rangle$ are the averages of
$u_{1x_1x_1}$ and $u_{2x_2x_2}$ conditional that the heights and
velocities fields be $h_1,h_2,u_1$ and $u_2$ with a spatial
difference $x$. Now we are interested in writing an evolution
equation for the PDF`s of height and its gradients difference. We
change the variables $h_1, h_2, u_1$ and $u_2$ with
$u_1=u-\frac{\omega}{2}$, $u_2=u+\frac{\omega}{2}$, $
h_1=h-\frac{\xi}{2}$ and $h_2=h+\frac{\xi}{2}$. Integrating over
$u$ and $h$ the PDF of the height and height gradient difference
is obtained
\begin{eqnarray}\label{Pdelta master}
 &&P^\delta(\xi,\omega,x,t) \cr \nonumber \\ &=&\int dh du
P(h-\frac{\xi}{2},h+\frac{\xi}{2},u-\frac{\omega}{2},u+\frac{\omega}{2},x,t)
\end{eqnarray}

Finally using the eq.(18), the master equation can be written as,
\begin{eqnarray}
P_{\omega \omega t}^\delta&=&-2\alpha
P_{x\omega}^\delta-\alpha(\omega P_x^\delta)_{\omega\omega} +
2(k(0)-k(x))P_{\omega\omega\xi\xi}^\delta \cr \nonumber \\
&+& 2(k_{xx}(0) -k_{xx}(x))P_{\omega\omega\omega\omega}^\delta+
G_{\omega\omega}^\delta
\end{eqnarray}

where by considering the definition of $G$ in (21), $G^{\delta}$
would be
\begin{eqnarray}\label{Gdelta}
  G^\delta(\xi,\omega,x,t)=\int dh du G.
\end{eqnarray}

It is clear that the $G^\delta$, which is proportional to surface
tension $\nu$, makes the master equation unclosed. In appendix A,
it is proved that for finite $\sigma$ in the limit of $\nu
\rightarrow 0$, $G^\delta$ can be written in terms of the
quantities which are defined on singularities.
%%%%%%%%%%%%%%%%%%%%%%%%%%%%%%%%%%%%%%%%%%%%%%%%%%%%%%%%%%%%%%%%%%%%%%%%%%
\begin{figure}
\epsfxsize=7truecm\epsfbox{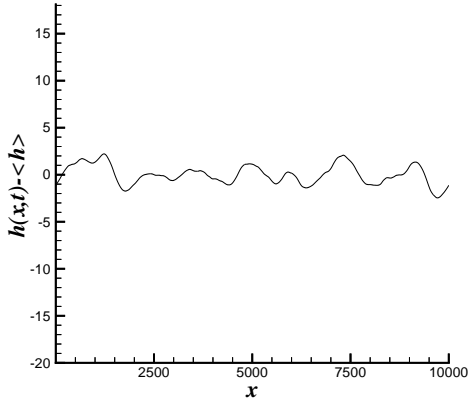}
\epsfxsize=7truecm\epsfbox{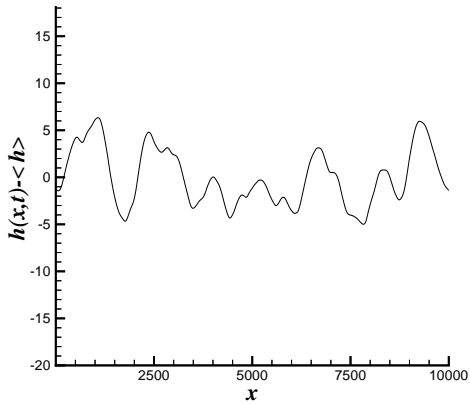}
\epsfxsize=7truecm\epsfbox{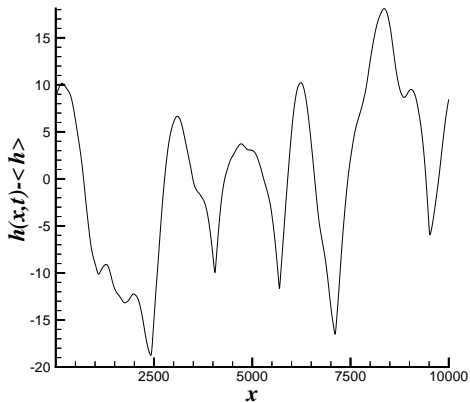} \narrowtext \caption{
 Different snapshots of the time-evolution of the
 height, with correlation lengths $\sigma \sim L/10 $ for the random
 periodic force, until the time that the system finally reaches to it's
stationary state. The average distance between the sharp valleys
is of order of $\sigma$ [56,57]. }
\end{figure}
%%%%%%%
%%%%%%%%%%%%%%%%%%%%%%%%%%%%%%%%%%%%%%%%%%%%%%%%%%%%%%%%%%%%%%%%%

%%%%%%%%%%%%%%%%%%%%%%%%%%%%%%%%%%%%%%%%%%%%%%%%%%%%%%%%%%%%%%%%%%%%%%%%%%
\begin{figure}
\epsfxsize=7truecm\epsfbox{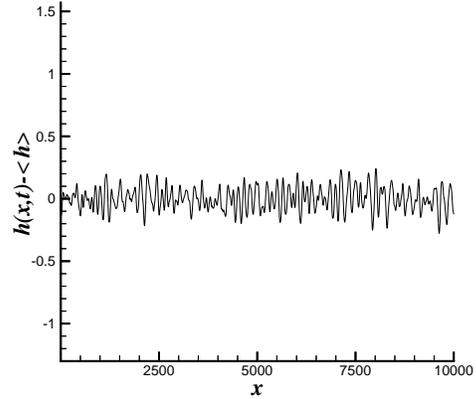}
\epsfxsize=7truecm\epsfbox{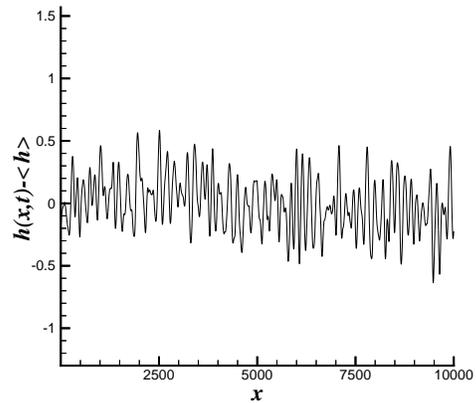}
\epsfxsize=7truecm\epsfbox{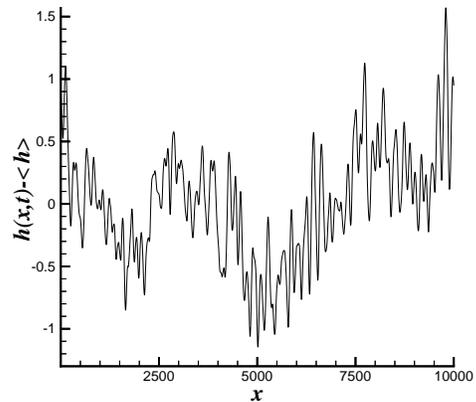}
 \narrowtext \caption{  Different snapshots of the time-evolution of the
 height, with correlation lengths $\sigma \sim L/100 $ for the random
 periodic force, until the time that the system finally reaches to it's
stationary state. }
\end{figure}
%%%%%%%
%%%%%%%%%%%%%%%%%%%%%%%%%%%%%%%%%%%%%%%%%%%%%%%%%%%%%%%%%%%%%%%%%

\section{Calculation of the moments and numerical simulation}

%%%%%%%%%%%%%%%%%%%%%%%%%%%%%%%%%%%%%%%%%%%%%%%%%%%%%%%%%%%%%%%%%%%%%%%%%%
\begin{figure}
\epsfxsize=7truecm\epsfbox{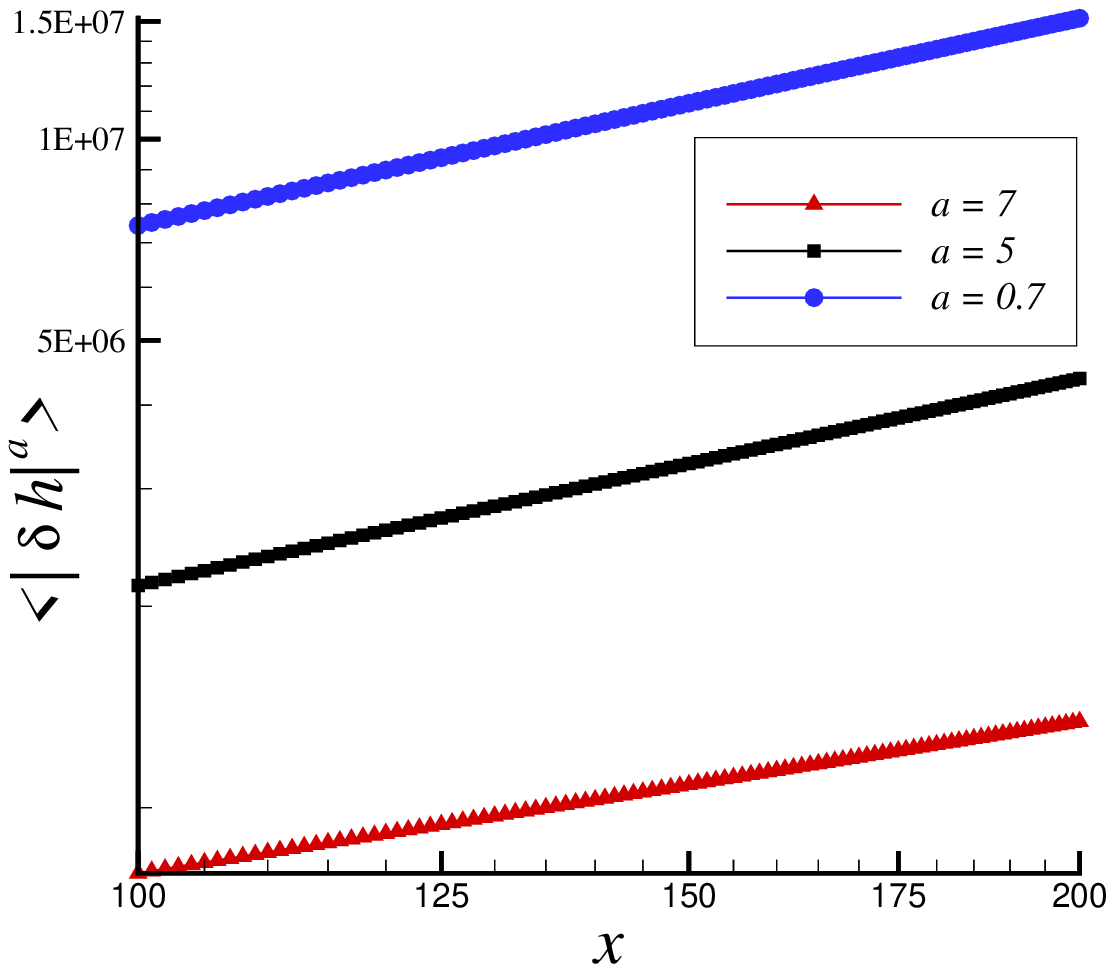}
\epsfxsize=7truecm\epsfbox{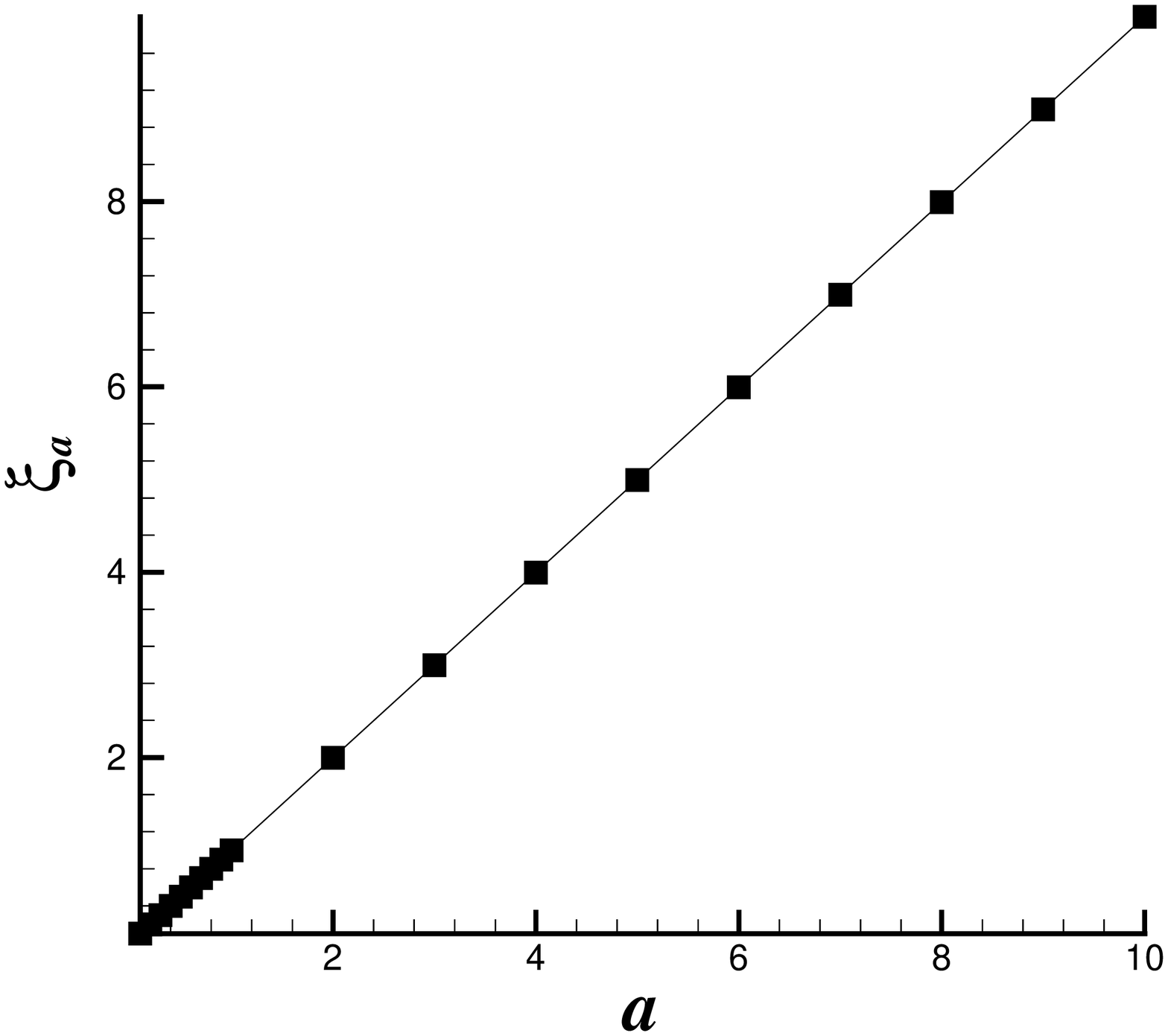}
\epsfxsize=7truecm\epsfbox{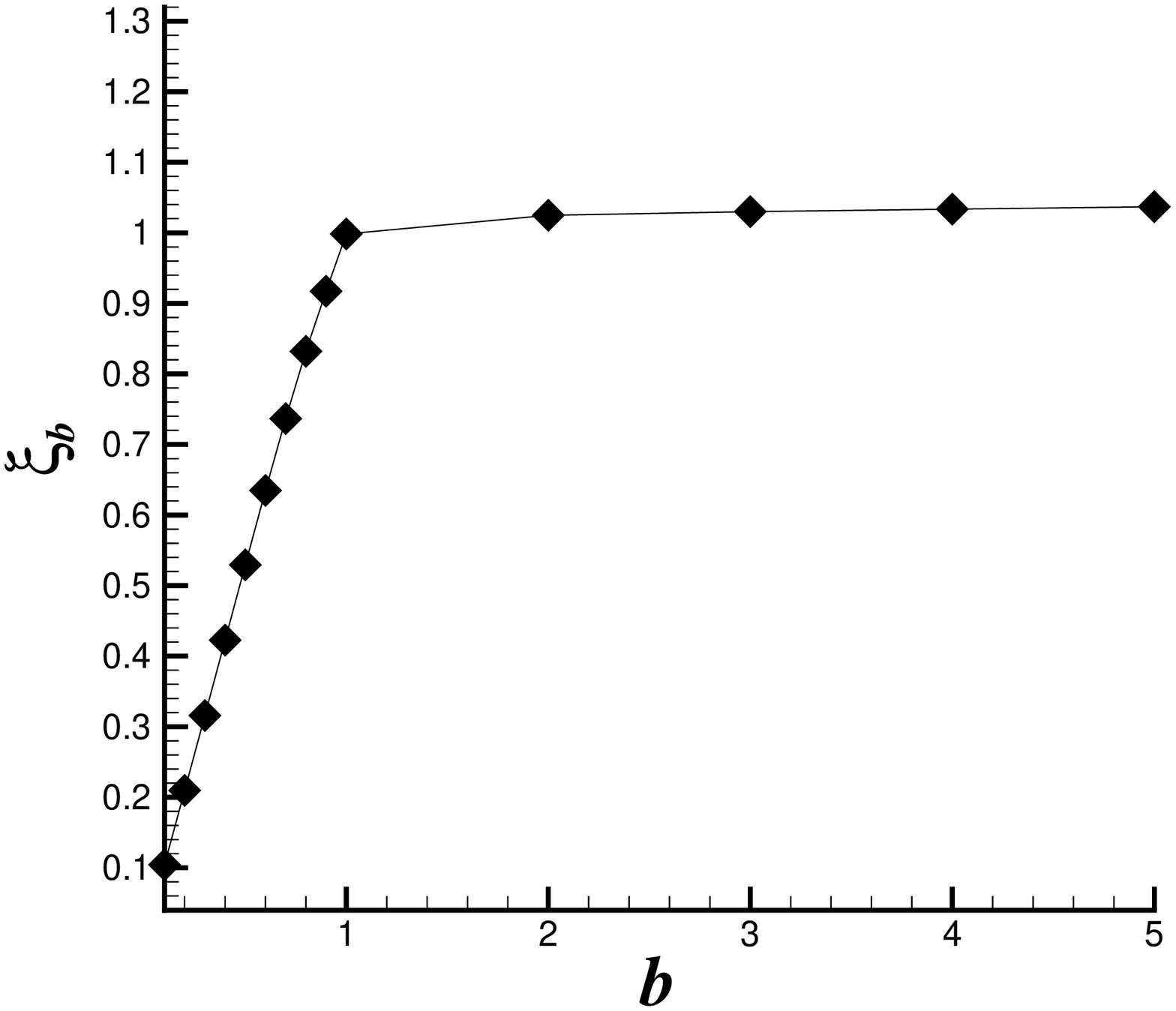}
 \narrowtext \caption{ In the upper graph the log-log plot of $ <
 |h(0) - h(x)|^a >$ vs $x$, for moments $a=0.7, 5$ and $7$ are demonstrated. In the middle
 figure the corresponding scaling exponent $\xi_a$ for
 height increments are plotted.
 The $\xi_a$ has a linear dependence on $a$. In the lower figure
 the scaling exponent $\xi_b$ for the moments of the height
 gradients increments
 are shown.}
\end{figure}
%%%%%%%%%%%%%%%%%%%%%%%%%%%%%%%%%%%%%%%%%%%%%%%%%%%%%%%%%%%%%%%%%

As shown in the previous section, the presence of surface tension
$\nu$, makes the master equation unclosed. However in the limit
$\nu \rightarrow 0$, that is the KPZ equation with an
infinitesimal surface tension, one can find exact scaling
exponents of the moments $\langle |h(x_1)- h(x_2)|^a
|u(x_1)-u(x_2)|^{b}\rangle$. It should be noted that the
$u$-field satisfying the Burgers equation,  for finite $\sigma$'s,
 develops discontinues or shock solutions
in the limit $\nu \rightarrow 0$. Consequently for finite $\sigma$
the height field shows up as a set of sharp valleys at the
positions where the shocks are located,
 continuously connected by some hill configurations,
as indicated in Figs.(1) and (2). As mentioned, each sharp valley
in position $y_0$ is identified by three quantities, namely the
gradients of $h$ in positions $y_{0^+}$, $ y_{0^-}$ and it's
height from $\bar h$. It is evident that the term $\lim _{\nu
\rightarrow 0} \nu u_{xx}$ is zero at the positions where no
sharp valley exists. Therefore in the limit $\nu\rightarrow 0$,
only small intervals around the sharp valleys will contribute to
the integral in the eq.(24).
 Within
these intervals, boundary layer analysis can be used for
obtaining an accurate approximation of $u(x,t), {\tilde h}(x,t)=
h - \bar h $. Generally, boundary layer analysis deals with those
problems in which perturbations are operative over very narrow
regions where the dependent variables undergo very rapid changes
across them. These narrow regions (sharp valley layers)
frequently adjoin the boundaries of the domain of interest, owing
the fact that a small parameter ($\nu$ in the present problem)
multiplies the highest derivative. A powerful method for treating
the boundary layer problems is the method of matched asymptotic
expansions. The basic idea underlying this method is that an
approximate solution to a given problem is sought not as a single
expansion in terms of a single scale, but as two or more separate
expansions in terms of two or more scales each of which is valid
in part of the domain. The scales are chosen, so that the
expansion as a whole, covers the whole domain of interest and the
domains  of validity of neighboring expansions overlap. In order
to handle the rapid variations in the sharp valley layers, a
suitable magnified or stretched scale and expand the functions in
terms of it in the sharp valley regions is defined. For this
purpose, we split $u$ and $\tilde h$ into a sum of inner solution
near the sharp valleys and an outer solution away from the sharp
valleys, and use systematic matched asymptotics to construct
uniform approximation of $u$ and $\tilde h$. It should be
emphasized that at point $y_0$ the height itself is continuous
and height gradient (corresponding Burgers velocity) is not
continuous.
 At these singular points the meaning of
$u_{\pm}$ is that $u_{\pm}(y_0,t)=u(y_{0 \pm },t)$.  Keeping in
mind that $u_{-}>u_{+}$,  the shock strength $s$ and the shock
velocity $\overline{u}$ are defined as $s=u_{+}-u_{-}$ and
$\overline{u}=\frac{1}{2}(u_{+}+u_{-})$.

 In appendices A and B, using the boundary layer method and the master equation,
  we have
proved analytically, that the joint moments of the height and the
corresponding gradient difference for any $a\geq 0 $ will be
\begin{eqnarray}\label{inequality}
   && \langle|\delta h|^a|\delta
u|^{b}\rangle \cr \nonumber \\ &=& \left\{
    \begin{array}{ll}
 {|x|^{a+b} \langle |\eta_h|^a|\eta_u|^b\rangle\quad}& \mbox{if}\
\ 0\le b<1,
      \\[6pt]
       {|x|^{a+1} \bigl( \langle |\eta_h|^a|\eta_u|\rangle+\frac{1}{2}\rho \langle|s|( |u_+|^a+|u_-|^a)\rangle\bigr)
        \quad}&\mbox{if}\ \ b=1
        \\[6pt]
{|x|^{a+1} \frac{1}{2}\rho\langle |s|^b( |u_+|^a+|u_-|^a)\rangle
        \quad}&\mbox{if}\ \ 1 < b
    \end{array}\right\}
\end{eqnarray}

where $\delta h = h(x_1) - h(x_2)$, $\delta u = u(x_1) - u(x_2)$
and $x=x_1-x_2$. The quantities $\eta_h$ and $\eta_u$ are the
regular parts of $\partial_x h $ and $\partial_x u $,
respectively. For $a=0$, our result will recover the known
results for Burgers equation with infinitesimal viscosity [47].

To prove the eq.(25), we have used the fact that the length scale
$\sigma$ is finite and $x$ is let to approach zero. This means
that we are dealing with  the  scaling  behaviour  of  the moments
 $ \langle|\delta h|^a|\delta u|^{b}\rangle$
for length scales $|x_1-x_2| << \sigma$.
 It is evident
that in these length scales the height increments fluctuations
are not intermittent. Indeed we find $\xi_a = a$ for any moments.

The moments  of  the   height  and  height-gradient increments
i.e.  $\langle|h( x_1 )-h(x_2 )|^{a}\rangle$ and
$\langle|u(x_1)-u(x_2)|^{b}\rangle$ are also calculated
numerically as a function of $ |x| = |x_1-x_2| $ for different
$a$'s and $b$'s. To simulate the problem , the KPZ equation is
discretized in space and time with scales $\delta$ and $dt$
respectively. The time scale $dt$ is related to $\delta$ as $dt
=\frac{3}{4}\frac{\delta}{u_m}$, where $u_m$ is the maximum of
the height gradient in each time step [78,79]. At each time step
the difference $U_j = |u(x_j+1)-u(x_j)|$ is checked for every
point $j$'s. For $U_j$`s that
 $ U_j > |\delta|^\frac{1}{3}$, we can determine the positions that the height field develop a sharp
 valley [57].
 Indeed this is a criterion for creation of a sharp valley in position $x_j=y_j$.
 At points that $ U_j < |\delta|^\frac{1}{3}$
the fields $u(x_j)$ and $h(x_j)$ belong to the smooth part.
Therefore the height fields $h(x_j)$ will fall into two regimes,
points far from the sharp valleys points $y_j$ and the points in
it's neighborhoods.
 For the points which the height field is regular or smooth, the height fields and it's corresponding
 gradients evolve under
 the KPZ and Burgers equation by setting the surface tension
 zero. Otherwise it is in the singularity or sharp valley region.
As mentioned in the introduction, every sharp valley can be
characterized by four parameters $s$, $\bar u$, $y_j$ and
$h(y_j)$. The time evolution of these quantities are given by the
following equations [77]
\begin{eqnarray}
&&\frac{dy_j}{dt}=\alpha {\bar u} \cr \nonumber\\
&&\frac{d}{dt} {u}(y_j)=\frac{\alpha}{4}s(h_{+xx}-h_{-xx})-f_x \cr \nonumber\\
&&\frac{d}{dt} s(y_j)=\frac{\alpha}{2}s(h_{+xx}+h_{-xx}) \cr
\nonumber \\
&& \frac{d}{dt}{\tilde h}(y_j,t) =-\frac{\alpha}{8}(4{\bar
u}^2-s^2) + f - \gamma
\end{eqnarray}
where $\gamma = {\bar h}_t$.

To calculate numerically the scaling behaviour of moments with $
x$ when $ x << \sigma$,  a
 periodic one dimensional substrate consisting a discrete N-point height
 field with the length of 10000 is used. Starting with a
 flat initial condition the height and its gradient fields evolve in time.
 We consider the random force
  as a white in time, smooth and periodic in space random function which it's
spatial correlation length is of the order of period of the given
periodic function. To generate this type of forcing we use the
kicking method which recently has been used in [51], to simulate
the Burgers turbulence. The basic idea is that the random force
can be decomposed as follows,

\begin{equation}
f(x,t) = \sum_j f_j(x)\,\delta(t-t_j) \label{kickforce}
\end{equation}
where $\delta$ is the Dirac distribution and where both the
``impulses'' $f_j(x)$ and the ``kicking times'' $t_j$ are
prescribed (deterministic or random). The kicking times are
ordered and form a finite or infinite sequence. In this article
the impulses are always taken smooth and acting only at length
scales $\sigma$. Newman \& McKane [80]  have used similar kicking,
in a context where the forced Burgers equation is used for the
study of directed polymers. Kraichnan [48] has considered a simple
model in which there are non-smooth impulses creating directly
saw-tooth profiles in the velocity in Burgers turbulence. Here
the time intervals are equal to the time steps of the algorithm's
run.

 In Figs.(1) and (2), we illustrate different snapshots of the time-evolution of the
 height, considering different correlation lengths $\sigma$ for the random
 periodic force, until the time that the system finally reaches to it's
stationary state. The following type of kicking force is used
\begin{eqnarray}
F(x,t)=A(t)[\cos(k x-\varphi(t))+\frac{1}{3}\sin(k x-\varphi(t))],
\end{eqnarray}
where $A$ is a white Gaussian random variable in time, which is
the noise amplitude and $\varphi$ is a homogeneous random phase.
Choosing  different values for $k$, leads to different values for
$\sigma$. The length scale $\sigma$ is of the order of the period
of $F$. In Fig.(3), the log-log plot of the moments of height
increments are sketched numerically for $a= 0.7, 5$ and $7$,
respectively.  We have found the exponents $\xi_a = r
 a + q$, where $r = 1.00 \pm 0.01$ and $q= - 0.0012 \pm 0.0002$.
 The scaling behaviour of the moments of height gradients increments for length scales $ x <<
\sigma$ is also checked. The results implies that with a good
precision $\langle |\delta u|^{b}\rangle$ scales with $x$ with
exponent $1$ for $b$'s larger than one, and scales with $x$ with
exponents $\xi_b = b$, for $b$'s smaller than one with precision
$\pm 0.001$. The behaviour of $\xi_b$ vs $b$ is also plotted in
fig.(3).

In summary, we study the problem of non-equilibrium surface
growth described by the forced KPZ equation in 1$+$1 dimensions.
The forcing is a white in time Gaussian noise but with a Gaussian
correlation in space with variance $\sigma$. Modeling a short
range correlated noise, we restrict our study to the case when
the correlation length of the forcing is much smaller than the
system size. Investigating the stationary state, a general
expression of the mixed correlations of height-difference and
height-gradient difference at any order, in terms of the length
scale $|x_1-x_2|$ and quantities which characterize the sharp
valley singular structures is given. Through a careful analysis
being done over the behaviour of the sharp valley environment, we
decipher the intermittency exponent of an arbitrary $a$-th
moment, i.e. $\langle |h(x_1)- h(x_2)|^a\rangle$. It is proved
that the height increments fluctuations are not intermittent and
its $a$-th moments for length scales  $|x_1-x_2| < \sigma$ scales
as $|x_1-x_2|^{\xi_a}$, where $\xi_a=a$. In the present paper the
limiting of $\nu\rightarrow 0$ is taken into account only for
finite $\sigma$ . Still the forcing correlation length is much
smaller than the system size and height correlation length. But
the limit $\sigma\rightarrow 0$ is a singular limit in our
calculations, and moreover, it is not a priori clear that the
limits of $\nu\rightarrow 0$ and $\sigma\rightarrow 0$ commute at
all. Using  stochastic equations which are governed over the
dynamics of quantities characterizing the sharp valleys we
simulate directly the problem and check the exponents. We have
generate the forcing using the kicking method. Our simulation
confirm the analytical results.  We believe that the analysis
followed in this paper is quite suitable for the zero temperature
limit in the problem of directed polymer in the random potential
with short range correlations [81]. The same method applied to KPZ
equation in higher dimensions would be definitely one of the
consequent goals of the present work. The main message which
might be encoded in the present work is the importance of the
statistical properties of the geometrical singular structures for
understanding the strong
coupling regime of Kardar-Parisi-Zhang equation.\\

\vskip +1cm

{\bf Acknowledgement}\\

We thank  A. Aghamohammadi F. Azami, M. Fazeli, F. Ghasemi and F.
Shahbazi
 for
their useful discussions.

\begin{center}
 { \bf \large Appendix A}
\end{center}

In this appendix we are going to prove that the $G$-term in
eq.(24), has a finite value in the limit $\nu \rightarrow 0$. As
shown in section 2 the $G$-term can be written as
\begin{eqnarray}
G=G(h_1,u_1,h_2,u_2,x,t)+G(h_2,u_2,h_1,u_1,-x,t).
\end{eqnarray}

Here we prove that in the vanishing surface tension limit, the $G$
term can be written as
\begin{eqnarray}
 &&G(h_1,u_1,h_2,u_2,x,t)= \cr \nonumber \\ && \rho \biggl( \int_{-\infty}^0\!\! d s \ s
  \int_{u_1+s/2}^{u_1-s/2}\!\! d\bar u \ (u_1-\bar u)T(\bar h_1,\bar
  u,s,h_2,u_2,x,t) \biggr)_{u_1}
\end{eqnarray}
where $T(\bar h_1,\bar
  u,s,h_2,u_2,x,t) $ is the PDF of
\begin{eqnarray}
  \nonumber (\bar h_1,\bar u(y_0,t),s(y_0,t),h_2(y_0+x,t),u_2(y_0+x,t))
\end{eqnarray}
conditional on $y_0$ being a sharp valley position.

Let us now prove the eq.(30). Assuming spatial ergodicity, for
example the average of one of the terms in $G$, which is
proportional to $ \nu$, can be expressed as
\begin{eqnarray}
  \label{eq:3.12}
 &&\nu \langle  u_{ix_ix_i}|h_1,h_2,u_1,u_2,x\rangle P =
  \nu \langle u_{ix_ix_i}(x,t) \delta(u_1-u_1(x_1,t))
   \cr \nonumber \\ && \delta(u_2-u_2(x_2,t))\delta(h_1- h_1(x_1,t))
  \delta(h_2- h_2(x_2,t)\rangle\nonumber\\
  &&=\nu\lim_{L\to\infty}\frac{N}{L}\frac{1}{N}\int_{-L/2}^{L/2} dx_i
  u_{ix_ix_i}(x_i,t)
   \delta(u_i-u_i(x_i,t))\cr \nonumber \\ && \hskip 3 cm
   \delta(h_i- h_i(x_i,t)).
 \end{eqnarray}

Clearly, in the limit as $\nu \to 0$ only small intervals around
the sharp valleys  will contribute to the integral. In these
intervals, boundary layer analysis can be used to obtain an
accurate approximation of $u_i(x,t)$ and $h_i(x,t)$. The basic
idea is to split $u_i$ and $h_i$ into the sum of an inner
solution near the sharp valleys and an outer solution away from
the singular point, and using systematic matched asymptotic to
construct uniform approximation of $u_1$ and $h_i$ (for details
see, e.g., [77]). For the outer solution, we look for an
approximation in the form of a series in $\nu$
\begin{eqnarray}
   h_i=h_i^{out}= h_i^0+\nu h_i^1+O(\nu^2)\nonumber\\
   u_i=u_i^{out}= u_i^0+\nu u_i^1+O(\nu^2).\nonumber
\end{eqnarray}

Then $u_i^0$ and $h_i^0$ satisfy
\begin{eqnarray}
  {h_i^0}_t-\frac{\alpha}{2}(\partial_{x_i}h_i^0)^2=f\nonumber\\
  {u_i^0}_t+\alpha u_i^0 {u_i^0}_{x_i}=-f_{x_i}
\end{eqnarray}
i.e. Burgers and KPZ equations without the surface tension terms.
In order to deal with the inner solution around the singularity,
let $y_i=y_i(t)$ be the position of a shock, and define the
stretched variable $z_i=(x_i-y_i)/\nu$ and let
\begin{eqnarray}
  \nonumber u_i^{in}(x,t)=v_i\left(\frac{x_i-y_i}{\nu}+\delta,t\right)
\end{eqnarray}
where $\delta$ is a perturbation of the sharp valley position to
be determined later.  Then, $v_i$ satisfies
\begin{eqnarray}
  \label{eq:3.13}
  \nu v_{it} +\alpha(v_i-\bar u_i+\nu \gamma)v_{iz} =v_{iz_iz_i}+\nu f
\end{eqnarray}
where $\bar u_i =dy_i/dt$, $\gamma = d\delta/dt$ and, to
$O(\nu^2)$, $\nu f$ can be evaluated at $x_i=y_i$ and can thus be
considered as a function of $t$ only.

We study eq.(\ref{eq:3.13}) by regular perturbation analysis. We
look for a solution in the form
\begin{eqnarray}
  \nonumber v_i= v_i^0 +\nu v_i^1+O(\nu^2).
\end{eqnarray}

To leading order, from eq.(\ref{eq:3.13}) we get for $v_i^0$ the
following equation
\begin{eqnarray}\label{alpha}
  \alpha(v_i^0-\bar u_i){v_i^0}_{z_i} ={v_i^0}_{z_iz_i}.
\end{eqnarray}
The boundary condition for this equation arises from the matching
condition with $u_i^{out}=u_i^0+\nu u_i^1 +O(\nu^2)$:
\begin{eqnarray}
  \nonumber \lim_{z_i\rightarrow \pm\infty} v_i^0 = \lim_{x_i\to y_i} u_i^0
  \equiv \bar u_i\pm\frac{s_i}{2}
\end{eqnarray}
where $s_i=s_i(t)$ is the sharp valley strength.  It is
understood that for small $\nu$ matching takes place for small
values of $|x_i-y_i|$ and large values of $|z_i|=|x_i-y_i|/\nu$.
This gives
\begin{eqnarray}
  \nonumber v_i^0= {\bar u_i}-\frac{s_i}{2} \tanh \left(\frac{s_i z_i}{4}\right).
\end{eqnarray}

These results show that, to $O(\nu)$, eq.(\ref{eq:3.12}) can be
estimated as
\begin{eqnarray}
 &&\nu \langle u_{ix_ix_i}|h_1,h_2,u_1,u_2,x\rangle P \cr \nonumber\\
  &&=\nu\lim_{L\to\infty}\frac{N}{L}\frac{1}{N}\sum_{i} \int_{\Omega_i} dx_i u_{ix_ix_i}^{in}(x_i,t)
  \delta(u_i-u_i^{in}(x_i,t))\cr \nonumber \\ &&
   \delta(h_i - h_i ^{in}(y_i,t)) \cr \nonumber\\
  &&=\nu \lim_{L \to \infty}\frac{N}{L}\frac{1}{N}\sum_{i}
\int_{-\infty}^{\infty} dz_i u_{iz_iz_i}^{in}
   \delta(u_i-u_i^{in}(z_i,t)) \cr \nonumber \\ &&
   \delta(h_i- h_i ^{in}(y_i,t))\nonumber\\
 &&= \nu\lim_{L\to\infty}\frac{N}{L}\frac{1}{N}\sum_{i}
\int_{-\infty}^{\infty} dz_i v_i^0{iz_iz_i}^{in}
   \delta(u_i-v_i^0)\cr \nonumber \\ &&
   \delta(h_i- h_i ^{in}(y_i,t))
\end{eqnarray}
where $\Omega_i$ is a layer centered at $y_i$ with width $\gg
O(\nu)$. Going to the stretched variable $z_i=(x_i-y_i)/\nu$,
 and using the eq.(\ref{alpha}), we have
\begin{eqnarray}
  \nonumber dz {v_0}_{zz}=dv_0\frac{{v_0}_{zz}}{{v_0}_z}= \alpha dv_0
  (v_0-\bar u)
\end{eqnarray}
so by taking the limit as $L\rightarrow \infty$, the $z $ integral
can be evaluated exactly
\begin{eqnarray}\nonumber
 &&\nu \langle
u_{1x_1x_1}|h_1,h_2,u_1,u_2,x\rangle P\nonumber\\
&&=\alpha\rho \int d\bar u
  \int_{-\infty}^{0}\!\!ds \ T(\widetilde{h_1},\bar u_1,s_1,\widetilde{h_2},u_2,x;t)
\cr \nonumber \\ &&
  \int_{\bar u_1+s_1/2}^{\bar
    u_1-s_1/2} \!\! dv_1^0\ (v_1^0-\bar u_1) \delta(u_1-v_1^0).
\end{eqnarray}
Where $({h_1},\bar u_1,s_1,{h_2},u_2,x;t)$ is the PDF of
$({h_1}(y_1,t),\bar
u_1(y_1,t),s(y_1,t),{h_2}(y_1+x,t),u_2(y_1+x,t))$ conditional on
$y_1$ being a sharp valley location and the spatial difference of
the heights $h_1$ and $h_2$ be $x$. Hence,
\begin{eqnarray}
  && \nu \langle
u_{1x_1x_1}|h_1,h_2,u_1,u_2,x\rangle P\nonumber\\
&&=-\alpha\rho\int_{-\infty}^0\!\! d s \ s
  \int_{u_1+s/2}^{u_1-s/2}\!\! d\bar u \ (u_1-\bar u)T(\bar h_1,\bar
  u,s,h_2,u_2,x,t).
\end{eqnarray}

For late use we note that the $ G $-term can be written in a more
convenient manner as
\begin{eqnarray}
&&G(h_1,u_1,h_2,u_2,x,t)=\nonumber\\
&&\frac{\rho}{2}  \int_{-\infty}^0\!\! d s \ s
  \big(T(\bar h_1,u_1-\frac{s}{2},s,h_2,u_2,x,t) \cr \nonumber \\ && +
   T(\bar h_1,u_1+\frac{s}{2},s,h_2,u_2,x,t)\big)\nonumber\\
 &&+ \rho\int_{-\frac{1}{2}}^{\frac{1}{2}}
  d\beta\int_{-\infty}^0\!\! d s \ s
  T(\bar h_1,u_1+\beta s,s,h_2,u_2,x,t).
\end{eqnarray}

\begin{center}
 {\bf \large Appendix B}
\end{center}

The main aim of this appendix is to calculate the mixed moments
$\langle|\delta h|^a|\delta u|^{b}\rangle$ by the use of master
equation derived in the section 2. As we will see the term $G$ has
an essential role in the results being to obtain the moments.
Considering in mind equations eqs.(21) and (22),
  $G^{\delta}$  could be written as
\begin{eqnarray}
 &&G^{\delta}(\xi,\omega,x,t)= \cr \nonumber \\ &&\int dhdu
G(h-\frac{\xi}{2},h+\frac{\xi}{2},u-\frac{\omega}{2},u+\frac{\omega}{2},x,t) \cr \nonumber\\
&&+  \int dhdu
G(h+\frac{\xi}{2},h-\frac{\xi}{2},u+\frac{\omega}{2},u-\frac{\omega}{2},-x,t).
\end{eqnarray}

It is proved in appendix A that the $G$-term can be written as
follows
\begin{eqnarray}
&&G(h_1,u_1,h_2,u_2,x,t)=
  \cr \nonumber \\ &&\alpha\rho \biggl( \int_{-\infty}^0\!\! d s \ s
  \int_{u_1+s/2}^{u_1-s/2}\!\! d\bar u \ (u_1-\bar u)T(\bar h_1,\bar
  u,s,h_2,u_2,x,t) \biggr)_{u_1}\nonumber\\
  \end{eqnarray}
where $T(\bar h_1,\bar
  u,s,h_2,u_2,x,t) $ is the PDF of
\begin{eqnarray} \nonumber
({\bar h_1},{\bar u(y_0,t)},s(y_0,t),h_2(y_0+x,t),u_2(y_0+x,t))
\end{eqnarray}
conditional on $y_0$ being a sharp valley position. It should be
emphasized that when we say $y_0$ is a singular point, we mean
that however the height itself is continuous at $y_0$ the height
gradient (corresponding Burgers velocity) is not continues at
these points. At these singular points the meaning of $u_{\pm}$ is
that $u_{\pm}(y_0,t)=u(\pm x,t))$ keeping in mind that
$u_{-}>u_{+}$, while the singularity strength $s$ and
$\overline{u}$ are defined as $s=u_{+}-u_{-}$ and
$\overline{u}=\frac{1}{2}(u_{+}+u_{-})$. We define $h_{+}(y_0,t)$
and $h_{+}(y_0,t)$ as
\begin{eqnarray}
h_{+}(y_0,t)=\overline{h}(y_0)+\frac{\epsilon}{2}\nonumber\\
h_{-}(y_0,t)=\overline{h}(y_0)-\frac{\epsilon}{2}
\end{eqnarray}

Due to the continuity of $h$ the limit $\epsilon\rightarrow 0$ is
not singular. Now let us rewrite the $G^{\delta}$ in a manner to
be more convenient for the rest of the calculations. For this
purpose let
\begin{eqnarray}\label{delu}
&&\delta u_{+}(x,y_0,t)=u(y_0+|x|,t)-u_{+}(y_0,t),\cr \nonumber &&
\delta u_{-}(x,y_0,t)=u_{-}(y_0,t)-u(y_0-|x|,t) \cr \nonumber
&&\delta h_{+}(x,y_0,t)=h(y_0+|x|,t)-h_{+}(y_0,t),\cr \nonumber
&& \delta h_{-}(x,y_0,t)=h_{-}(y_0,t)-h(y_0-|x|,t) \cr \nonumber
\end{eqnarray}
 and define $U_{\pm}(\epsilon,s,\delta h_{\pm},\delta u_{\pm},x,t)$ be the
PDF's of $(\epsilon,s(y_0,t),\delta h_{\pm}(x,y_0,t),\delta
u_{\pm}(x,y_0,t))$ conditional on $y_0$ being a sharp valley
position. Then $G^{\delta}$ can be expressed as
\begin{eqnarray}
G^{\delta}(\xi,\omega,x,t)=G_{+}^{\delta}(\xi,\omega,x,t)+G_{-}^{\delta}(\xi,\omega,x,t)
\end{eqnarray}
where
\begin{eqnarray}\label{Gplusmin}
&&G_{\pm}^{\delta}(\xi,\omega,x,t)= \cr \nonumber\\
&& \alpha\frac{\rho}{2} \int_{-\infty}^{0} ds
s[U_{\pm}(\epsilon,s,sgn(x)
\xi-\epsilon,sgn(x)\omega-s,x,t) \cr \nonumber \\ && +U_{\pm}(\epsilon,s,sgn(x)\xi,sgn(x)\omega,x,t)]\cr \nonumber\\
&&-\alpha\rho \int_{-\infty}^{0}dss \int_{0}^{1}d\beta
U_{\pm}(\epsilon,s,sgn(x)\xi-\frac{\epsilon}{2},sgn(x)\omega-\beta
s ,x,t).
\end{eqnarray}

We are interested in scaling behaviour of mixed moments in small
length scale $x$. In the limit $x\rightarrow 0$ it should be noted
that $P^{\delta}$ can be decomposed into two parts as
\begin{eqnarray}
P^{\delta}(\xi,\omega,x,t) & = &
p_{ns}(x,t)P^{\delta}(\xi,\omega,x,t|no\hskip .1cm sharp \hskip
.1cm valley ) \cr \nonumber \\ &+&
(1-p_{ns}(x,t))P^{\delta}(\xi,\omega,x,t|sharp \hskip .1cm valley
)
\end{eqnarray}
where $p_{ns}(x,t)$ is the probability that there is no sharp
valley in $[y,y+x)$ and  $P^{\delta}(\xi,\omega,x,t|no \hskip .1cm
sharp \hskip .1cm valley )$ is the PDF of $\delta u(x,y,t)$ and
$\delta h(x,y,t)$ conditional on the property that there is no
sharp valley in $[y,y+x)$. Also  $P^{\delta}(\xi,\omega,x,t| sharp
\hskip .1cm  valley)$ is the PDF of $\delta u(x,y,t)$ and $\delta
h(x,y,t)$ conditional on the property that there is at least one
sharp valley in $[y,y+x)$. Since by definition of number density
of sharp valleys $\rho$ we have
\begin{eqnarray}
p_{ns}= 1-\rho|x|+o(x)
\end{eqnarray}
\begin{eqnarray}
P^{\delta}(\xi,\omega,x,t|sharp \hskip .1cm valley
)=R(\xi,\omega,x,t)+O(1)
\end{eqnarray}
where $R(\xi,s,x,t)$ is the PDF of $\xi=h(y_0+x)-h(y_0)$,
$s(y_0,t)$ and $x$, conditional that $y_0$ be a shock position.
\begin{eqnarray}
 && p_{ns}(x,t)P^{\delta}(\xi,\omega,x,t|no \hskip .1cm sharp \hskip
.1cm valley ) \cr \nonumber \\
&&
=(1-\rho|x|)\frac{1}{x^2}Q(\frac{\xi}{x},\frac{\omega}{x},t)+o(x)
\end{eqnarray}
here $Q(\eta_{h},\eta_{u},t)$ is the PDF of $\eta_{h}(x,t)$ and
$\eta_{u}(x,t)$, the regular part of the velocity and the velocity
gradient, respectively. Indeed we have considered the case $x>0$.
The case $x<0$ can be treated similarly. We note that, in the
limit $x \to 0$, because of dealing with
  regular points, we have
\begin{eqnarray}
  \nonumber x^2P^\delta(x \eta_h,x \eta_u,x,t)\to Q(\eta_h,\eta_u,t).
\end{eqnarray}

It implies that
\begin{eqnarray}
  \nonumber P^\delta(\xi,\omega,x,t)=
  \delta(\omega)\delta(\xi)+o(1).
\end{eqnarray}

 Define
\begin{eqnarray}
  \nonumber A(\xi,\omega,t) &=& \lim_{x\to0} x^{-1}(P^\delta(\xi,\omega,x,t)-\delta(\omega)\delta(\xi)) \cr \nonumber \\
  &=&
  \lim_{x\to0} {P}_x^\delta(\xi,\omega,x,t).
\end{eqnarray}

Taking the limit as $x\to0$ in the equation for $P^\delta$ (
eq.(23) ) and considering that the system has reached to the
stationary state, it follows that $A$ satisfies
\begin{eqnarray}
  \label{eq:3.a.1}
  0&=&-\alpha\omega A-2\alpha\int d\omega'\ H(\omega,-\omega') A(\xi,\omega',x,t)
\cr \nonumber \\
  &+&B(\xi,\omega,t)
\end{eqnarray}
where we have used $\lim_{x\to0}(K(0)-K(x))=0$ and also we defined
\begin{eqnarray}
  \nonumber B(\xi,\omega,t)=\lim_{x\to0}G^\delta(\xi,\omega,x,t).
\end{eqnarray}

To evaluate $B$ note that as $x\to0$
\begin{eqnarray}
  \nonumber \delta u_{\pm} (x,y_0,t)\to0.
\end{eqnarray}

 This implies that, as $x\to0$,
\begin{eqnarray}
  U_\pm(s,\xi,\omega,x,t) \to
  S(s,t)\delta(\omega)\delta(\xi)\nonumber\\
\end{eqnarray}
where $S(s,t)$ is the PDF of $s(y_0,t)$ conditional on $y_0$ being
a sharp valley location. Hence, from the expression for
$G^\delta$,
\begin{eqnarray}
 && \nonumber B(\xi,\omega,t)=\alpha\rho \omega S(\omega,t)\delta(\xi)+\alpha\rho < s>
   \delta(\omega)\delta(\xi)\cr \nonumber \\ &+& 2\alpha\rho\delta(\xi)
  \int_{-\infty}^\omega \!\! d\omega' S(\omega',t)- 2\alpha\rho H(\omega)\delta(\xi)
\end{eqnarray}
where $H(\cdot)$ is the Heaviside function and we used $S(s,t)=0$
for $s>0$ since $s(y_0,t)\le0$.  Inserting this expression in
(\ref{eq:3.a.1}), the solution of this equation is
\begin{eqnarray}
  \nonumber A(\xi,\omega,t)=(-\delta(\omega)+\rho \langle s\rangle \delta^1(\omega)+\rho
  S(\omega,t))\delta(\xi)
\end{eqnarray}

Here $\delta^1(\omega)=d\delta(\omega)/d\omega$ and $\omega$ used
the identity $\omega\delta^1(\omega)=-\delta(\omega)$. Using the
fact that $\rho<s>=-< \eta_u>$ [47], we can be restated
$A(\xi,\omega,t)$ as
\begin{eqnarray}
  \nonumber A(\xi,\omega,t)=(-\delta(\omega)-\langle\eta_u\rangle\delta^1(\omega)+\rho
  S(\omega,t))\delta(\xi).
\end{eqnarray}

Hence, combining the above results, we have
\begin{eqnarray}
  \nonumber P^\delta(\xi,\omega,x,t) &=&
 (\delta(\omega)-x(\delta(\omega)+\langle \eta_u\rangle\delta^1(\omega)- \rho
  S(\omega,t)))\delta(\xi)\cr \nonumber \\ &+& o(x).
\end{eqnarray}

 Which is correct for
$x>0$. We Reorganize this expression as
\begin{eqnarray}
   P^\delta(\xi,\omega,x,t) &=&[(1-\rho x)(\delta(\omega)-x\langle \eta_u\rangle \delta^1(\omega))+ x\rho
  S(\omega,t)]\delta(\xi)\cr \nonumber \\ &+& o(x)\nonumber
\end{eqnarray}
 and then we use the identity
\begin{eqnarray}
  \nonumber \delta(\omega)-x\langle \eta_u\rangle \delta^1(\omega)=\frac{1}{x^2}
  Q\left(\frac{\xi}{x},\frac{\omega}{x},t\right)+o(x).
\end{eqnarray}

 Now
we decompose the fields $h$ and $u$ in terms of the their regular
and singular parts as
\begin{eqnarray}
  \nonumber h_x(x,t)= \eta_{h}(x,t)+\sum_j \epsilon(y_j,t) \delta(y-y_j)
\end{eqnarray}
and
\begin{eqnarray}
  \nonumber u_x(x,t)= \eta_{u}(x,t)+\sum_j s(y_j,t) \delta
  (y-y_j).
\end{eqnarray}
 So if we let $f^\delta(\omega,x,t)$ be defined as
\begin{eqnarray}
  \nonumber f^\delta(\xi,\omega,x,t)&=(1-\rho |x|)\frac{1}{x}
  Q\left(\frac{\xi}{x},\frac{\omega}{x},t\right)
  +|x|\rho R(\xi,\omega,x,t)
\end{eqnarray}
 then we can write
 \begin{eqnarray}
P^{\delta}(\xi,\omega,x,t)&=&f^\delta(\xi,\omega,x,t)+o(x).
 \end{eqnarray}

Now we can prove eq.(25) for  $0 \le b \le1$ and an arbitrary
value of $a$. The proof for other values of $b$ is similar.  Let
\begin{eqnarray}
  \nonumber  &&f^\delta(\xi,\omega,x,t) \cr \nonumber \\ &=&(1-\rho |x|)\frac{1}{x^2}
  Q\left(\frac{\xi}{x},\frac{\omega}{x},t\right)
  +|x|\rho R(\xi,\omega,x,t),\cr \nonumber \\
  && g^\delta(\xi,\omega,x,t)= \cr \nonumber \\ &&[\delta(\omega)-|x|(\rho
  \delta(\omega)+\langle\xi\rangle
  \delta^1(\omega)-\rho S(\omega,t))]\delta(\xi) \nonumber
\end{eqnarray}

Because the sharp valley points have contribution in large
$\omega$'s we can write for $M > 0$
\begin{eqnarray}
  \nonumber
  &&\int d\xi d\omega |\xi|^a |\omega|^b (Z^\delta-f^\delta)\\
  &=& \int_{|\omega|\le M}\!\!d\xi d\omega |\xi|^a |\omega|^b (Z^\delta-f^\delta) \cr \nonumber \\ &+& \int_{|\omega|>
    M}\!\! d\xi d\omega |\xi|^a |\omega|^b  (Z^\delta-f^\delta).
\end{eqnarray}

Because of eq.(52), the first term at the rhs of eq.(54) is
$o(x)$.  To estimate the second term, note that for $M$ large
enough
\begin{eqnarray}
   && \int_{|\omega|> M}\!\! d\xi d\omega |\xi|^a |\omega|^bZ^\delta \le
  \int_{|\omega|> M}\!\! d\xi d\omega |\xi|^a \omega^2 Z^\delta\nonumber\\
  &\le& \left|\int d\xi d\omega |\xi|^a \omega^2 (Z^\delta-g^\delta)\right|+
  \int_{|\omega|> M}\!\! d\xi d\omega |\xi|^a \omega^2 g^\delta\nonumber\\
  &=&o(x) +|x|\rho \int_{|\omega|> M}\!\! d\xi d\omega |\xi|^a \omega^2 S(\omega,t)\delta(\xi)\nonumber\\
  &=&o(x).
\end{eqnarray}

Because the singular part of $Z$ is cancelled  by $g$, the first
term should be of order of $o(x)$
\begin{eqnarray}
  && \int_{|\omega|> M}\!\! d\xi d\omega |\xi|^a |\omega|^bf^\delta\cr \nonumber &=&|x|^{a+b} (1-|x|\rho)
  \int_{|\eta_u|> M/x}\!\! d\eta_h d\eta_u \ |\eta_h|^a \ |\eta_u|^b Q(\eta_h, \eta_u,t) \cr \nonumber
  &+ & |x|\rho\int_{\omega> M}\!\! d\xi d\omega |\xi|^a
  |\omega|^bR(\xi,\omega,x,t)
    =o(x^{a+b}).
\end{eqnarray}

We can write  $R(\xi,\omega,x,t)$ as
$\frac{1}{2}R(\xi_+,\omega,|x|,t)+\frac{1}{2}R(\xi_-,\omega,-|x|,t)$
 where for $R(\xi_+,\omega,|x|,t)$ and $R(\xi_-,\omega,|x|,t)$, we
have the condition that $\xi=h(y_0+|x|)-h(y_0)$ and
$\xi=h(y_0+|x|)-h(y_0)$, respectively. When $x\rightarrow 0$ we
can write $\xi_\pm =u_\pm |x|$.

Since $M$ can be made arbitrarily large, we get
\begin{eqnarray}
  \nonumber \int d\xi d\omega |\xi|^a |\omega|^b(Z^\delta-f^\delta)\le o(x^{a+b})+\delta_M O(x)
\end{eqnarray}
where $\delta_M\to0$ as $M\to+\infty$. Noting that
\begin{eqnarray}
   &&\int d\xi d\omega |\xi|^a |\omega|^bf^\delta \cr \nonumber \\ &=& \left\{
    \begin{array}{ll}
      {|x|^{a+b} \langle |\eta_h|^a|\eta_u|^b\rangle\quad}& \mbox{if}\ \ 0\le b<1
      \\[6pt]
      {|x|^{a+1} \bigl( \langle |\eta_h|^a|\eta_u|\rangle+\frac{1}{2}\rho \langle|s|( |u_+|^a+|u_-|^a)\rangle\bigr)
        \quad}&\mbox{if}\ \ b=1
    \end{array}\right.
\end{eqnarray}

We obtain (\ref{inequality}) for $0\le b\le 1$.  For $b
> 1$ the leading term in our calculation, will be the second term of the eq.(56) with the order of
$o(x^{a+1})$. The leading term is $\frac{1}{2}\rho \langle|s|(
|u_+|^a+|u_-|^a)\rangle$.

  Also there is an alternative method to prove the eq.(25) for the
case $b > 1$. The method is based on the calculation of the mixed
moment
 $\langle(|h(x_1)-h(x_2))|^a|(u(x_1)-u(x_2))|^b\rangle$ for integer orders while $b \geq 1$, directly from the
 PDF`s equation (23) by integrating  over two
 $\omega$'s, i.e.
\begin{eqnarray}\label{app master}
 P^\delta_t&=&-\alpha\omega P^\delta_{x}-2\alpha\int d\omega'\ H(\omega'-\omega)
  P^\delta_x(\xi,\omega',x,t)\cr \nonumber \\
  &&+2(K_{xx}(0)-K_{xx}(x)) P^\delta_{\omega\omega}+2(K(0)-K(x))
  P^\delta_{\xi\xi} \cr \nonumber \\ &+& G^\delta(\xi,\omega,x,t).
\end{eqnarray}

 In the limit $x\rightarrow 0$, keeping $\sigma$ finite, and in stationary state ,
it will be simplified to
\begin{eqnarray}
0&=&-\alpha\omega P^\delta_{x}-2\alpha\int d\omega'\
H(\omega'-\omega)
  P^\delta_x(\xi,\omega',x,t) \cr \nonumber \\ &+& G^\delta(\xi,\omega,x,t)
\end{eqnarray}

 First of all the term $\int d\xi d\omega |\xi|^n |\omega|^m G^{\delta}(\xi,\omega,x,t)$
should be  calculated in the $x\rightarrow\pm 0$ limit. This can
be done by using the relation (\ref{Gplusmin}). Note that

\begin{eqnarray}
 && \int d\xi d\omega |\xi|^n |\omega|^m
G^{\delta}(\xi,\omega,x,t)= \cr \nonumber\\
&&\alpha\frac{\rho}{2}\langle
  s|(\delta u_++{\rm sgn}(x) s)|^m|(\delta h_++{\rm sgn}(x)
  \epsilon)|^n\rangle \cr \nonumber \\
  &+& \alpha\frac{\rho}{2} \langle s|\delta u_+|^n|\delta
  h_+|^n\rangle \cr \nonumber\\
  &-&\alpha\rho \int_0^1 \!\! d\beta\
  \langle s|(\delta u_++\beta{\rm sgn}(x) s)|^m|(\delta h_++{\rm
  sgn}(x)\frac{\epsilon}{2})|^n\rangle \cr \nonumber\\
  &+&\alpha\frac{\rho}{2} \langle s|( \delta u_-+{\rm sgn}(x)s)|^m|( \delta h_-+{\rm sgn}(x)\epsilon)|^n\rangle
  \cr \nonumber \\
  &+& \alpha\frac{\rho}{2} \langle s|\delta u_-|^m|\delta
  h_-|^n\rangle \cr \nonumber\\
  &-&\alpha\rho \int_0^1 \!\! d\beta\ \langle s|(\delta u_-+\beta{\rm sgn}(x)
  s)|^m|(\delta h_-+{\rm sgn}(x)\frac{\epsilon}{2})|^n\rangle.
  \end{eqnarray}

  If we go back and look carefully to the definition $\delta u_{\pm}$, we
  see that
   $\delta h_{\pm}\simeq  u_{\pm}|x|\simeq o(x)$ and $\delta u_{\pm}\simeq
  o(x)$ as $x\rightarrow 0$.  While it should be realized that the sharp valley  strength $s$ is of the
  order $O(1)$ as $\epsilon\rightarrow 0$, so in the limit
  $x\rightarrow
  0$, the result of the integral would be simplified
  as
\begin{eqnarray}
 &&\int d\xi d\omega |\xi|^n |\omega|^m
G^{\delta}(\xi,\omega,x,t)\simeq \cr \nonumber \\ &&
\frac{\alpha}{2}(\frac{m-1}{m+1})\rho |x|^{n+1} \langle
|s|^{m+1}( |u_+|^n+|u_-|^n)\rangle.\nonumber
\end{eqnarray}

 Finally  multiplying the  terms of equation (\ref{app master}) in $|\xi|^n$ and $|\omega|^m$ and
 integrating respect to $\xi$ and $\omega$ we have
 \begin{eqnarray}\label{integer moments}
&&  \langle |\xi|^n|\omega|^{m+1}\rangle = \cr \nonumber \\ &&
\frac{1}{2}\rho |x|^{n+1} \langle |s|^{m+1}(
|u_+|^n+|u_-|^n)\rangle \hskip .5 cm if \hskip .5 cm \ m \geq
n\in {\cal N}
\end{eqnarray}
where the result coincides perfectly with eq.(\ref{inequality})
which is the general form of eq.(\ref{integer moments}).

\bibliographystyle{plain}

\end{document}